\documentclass[sigconf]{acmart}
\usepackage[ruled,vlined]{algorithm2e}
\usepackage{url}
\usepackage{caption}
\SetKwInput{KwInput}{Input}                
\SetKwInput{KwOutput}{Output}              

\AtBeginDocument{%
  \providecommand\BibTeX{{%
    \normalfont B\kern-0.5em{\scshape i\kern-0.25em b}\kern-0.8em\TeX}}}

\copyrightyear{2020}
\acmYear{2020}
\setcopyright{acmlicensed}\acmConference[ICAIF '20]{ACM International Conference on AI in Finance}{October 15--16, 2020}{New York, NY, USA}
\acmBooktitle{ACM International Conference on AI in Finance (ICAIF '20), October 15--16, 2020, New York, NY, USA}
\acmPrice{15.00}
\acmDOI{10.1145/3383455.3422556}
\acmISBN{978-1-4503-7584-9/20/10}

\begin{document}
\fancyhead{} 

\title{Social media data reveals signal for public consumer perceptions}

\author{Neeti Pokhriyal}
\affiliation{%
  \institution{Department of Computer Science \\Dartmouth College}}
\email{neeti.pokhriyal@dartmouth.edu}

\author{Abenezer Dara}
\affiliation{%
  \institution{Department of Computer Science \\Dartmouth College}}
\email{abenezer.d.dara.20@dartmouth.edu}

\author{Benjamin Valentino}
\affiliation{%
  \institution{Department of Government \\Dartmouth College}}
\email{benjamin.a.valentino@dartmouth.edu}


\author{Soroush Vosoughi}
\affiliation{%
  \institution{Department of Computer Science \\Dartmouth College}}
\email{Soroush.Vosoughi@dartmouth.edu}

\begin{abstract}
Researchers have used social media data to estimate various macroeconomic indicators about public behaviors, mostly as a way to reduce surveying costs. One of the most widely cited economic indicator is consumer confidence index (CCI). Numerous studies in the past have focused on using social media, especially Twitter data, to predict CCI. However, the strong correlations disappeared when those models were tested with newer data according to a recent comprehensive survey. In this work, we revisit this problem of assessing the true potential of using social media data to measure CCI, by proposing a robust non-parametric Bayesian modeling framework grounded in Gaussian Process Regression (which provides both an estimate and an uncertainty associated with it). Integral to our framework is a principled experimentation methodology that demonstrates how digital data can be employed to reduce the frequency of surveys, and thus periodic polling would be needed only to calibrate our model. Via extensive experimentation we show how the choice of different micro-decisions, such as the smoothing interval, various types of lags etc. have an important bearing on the results. By using decadal data (2008-2019) from Reddit, we show that both monthly and daily estimates of CCI can, indeed, be reliably estimated at least several months in advance, and that our model estimates are far superior to those generated by the existing methods. 


\end{abstract}


\maketitle

\section{Introduction}
Sample surveys have been the de facto instrument for understanding the social, political and economic realities of the population. Social scientists have long used these sample surveys to gather information from a subset of population and generalize their results to larger populations of interest. However, traditional surveys are time consuming and money intensive exercises. Also with dwindling public participation~\cite{nonresponse}, especially in phone surveys, researchers have begun exploring alternative data sources, like social media (SM) data to either supplant or supplement the traditional surveys. 

Using SM to construct official indicators has a number of advantages. First, these datasets are collected continuously and provide for more frequent sensing of public perceptions than administrative and polling data can provide. Such timely data might offer market analysts with almost real-time information, and assist in when economic decision must be made needed prior to the availability of official indicators. Second, SM are available at a lower (or no) cost compared to traditional polling, which often cost tens or hundreds of thousands of dollars to administer. Third, analyzing SM offers a unique opportunity to glean signals from personal conversations and community discussions occurring naturally, which might assist in capturing additional signs/parameters about the population, in contrast to the responses to a few numbered questions asked in a poll. Some researchers have even likened social media generated models of public opinion as ex-post “surveys"~\cite{laborflows}.

We focus on consumer confidence index (CCI), one of the most widely cited economic surveys in the United States. Accurately measuring consumer confidence is pivotal to governments, businesses, media and individuals as it describes both current economic conditions and captures consumer’s hope for near future~\cite{cci}, and influences whether the economy expands or contracts, as consumer spending drives 70\% of the GDP~\cite{ics}.

Researchers have argued that since SM provides a means to sense social attitudes and behaviors, it can act as a proxy for surveys that attempt to gauge consumer confidence, and proposed models to do that~\cite{connor2010,poq18,laborflows}. However, they have largely been inconclusive when tested with newer and longer data~\cite{sage19}.

In this work, we revisit the problem of accessing the true potential of SM data to predict CCI, by proposing a robust Bayesian framework grounded in Gaussian Process Regression. Central to our framework is a principled validation methodology that describes how SM data can be used to reduce the frequency of surveys. With extensive experimentation, we show how the choice of different micro-decisions, like smoothing interval, various types of lags etc have an important bearing on the inferences drawn. This analytic framework, along with the novel validation strategy to reduce survey frequency, is our main contribution. Additionally, we also provide baselines to our models to evaluate the practical utility of substituting predictions based on SM data for surveys.  

We show how accurate high frequency estimates of CCI can be reliably generated from SM, which are far superior to those generated by the baselines methods, namely autoregressive framework and a method based on SM sentiment. While we do not obviate the need for regular polls; we do demonstrate that by employing our method, polls can be conducted less frequently. We show how periodic polling would be needed only to calibrate our model, thus generating significant cost savings for governments and organizations that produce and use CCI.

We validate our predictions through extensive empirical testing using the longitudinal data from 2008 to 2019, and demonstrate the accuracy of our model on both daily and monthly aggregations of CCI. We show that CCI can be reliably estimated at least several months in advance, and that our model estimates capture macro and micro trends in the CCI.

The SM of our choice is Reddit, an online forum where of users post links and discuss their ideas and interests. As of Oct'19, Reddit had 430 million monthly active users with 199 million posts and 1.7 billion comments~\cite{reddit2019}. Our underlying hypothesis is that the user conversations recorded on Reddit forums related to health of the economy, buying and selling decisions of individuals, job prospects, financial opportunities etc., correlate with the underlying signal of consumer confidence.

While most previous works use Twitter for estimate CCI, we use Reddit data. There is an important difference in communicative dynamics between the two platforms - Twitter is more appropriate to study connections and diffusion of information~\cite{twitterorreddit}, whereas Reddit provides a discussion platform through which individuals can discuss their plans of buying commodities or houses, job losses or gains and general chatter about economy. Since CCI is ultimately intended to understand human behavior, we feel Reddit is more likely to capture public optimism or pessimism in economic affairs.

\section{Related Works}
Researchers have explored extracted sentiments from Twitter, Facebook posts, news etc. and their correlation with the CCI as measured via Gallup polls, Michigan surveys etc. with mixed results~\cite{connor2010,poq18,surveySM,sage19}. 

Most existing studies use a single measure (mostly a sentiment feature) and study correlations and use it to forecast CCI values. Most of these studies are based on the sentiment of few keywords, such as jobs on Twitter. These analyses utilize methods that look for correlations over a short period of time or with coarser granularity of data, e.g. monthly. Thus, it is important to study the generalizabilty of these models with longer time series. Existing works have also fail to assess the temporal stability of their estimates. This kind of analysis is critical, however, if we seek to evaluate whether SM can be used as a potential replacement or substitution for surveys~\cite{surveySM}. 
While most existing studies have used Twitter or Google query~\cite{googlequery} or purchase history data~\cite{purchase}, we use Reddit to understand CCI. Researchers have used Reddit data for a variety of related applications such as understanding online user behavior~\cite{abuse}, (dis)information propagation, health informatics~\cite{juul} etc, but to the best of our knowledge, Reddit has not been used to study public opinion on economic  issues.

Gaussian Processes have been studied extensively in geostatistics (as kriging), meterology etc., as they provide high flexibility in modeling as well as a principled way of learning the hyper-parameters~\cite{gpml}.

\section{Methodology}

Briefly, our computational framework for predicting CCI entails extracting quantitative metrics that capture the content, psychological as well as sentiment features from Reddit posts and comments. We also handle noise in the SM data by accessing trustworthy users. Later, the features are aggregated to daily and, further, monthly granularity, and used to build a Gaussian Process regression (GPR) model with targets as daily (and monthly) CSI. 

\subsection{Data Transformation}

Reddit is based on a user-to-topic subscription model, and is divided into "subreddits", which are content feeds focused on particular topics, such as finance or jobs, etc. User interaction occurs when users post something of interest (personal narrative, news etc.), and others comment on the post. We study the texts written on the relevant subreddits in the form of posts. Each post is specified by a timestamp and its author. Other users can either upvote, downvote or comment on previous posts. For each post, we extract three types of features based on content, sentiment and comments


\subsubsection{Content Extraction}:
We employ {\em doc2vec} embeddings, where each text is mapped into a vector such that semantically similar texts have more similar vector representations than dissimilar ones~\cite{doc2vec}. Though most of the previous research focus solely on sentiment features, we extract both the sentiments and text content of the posts. We argue that content features are important and empirically show that these features alone capture the overall trend of the consumer sentiment. The rationale behind using content features is that SM users generate posts on temporally relevant and important issues, since the main motivation for posting is to elicit responses from the community. This is especially true for economic variables, as SM users can be expected to write and reflect more about recent economic happenings when economic projections are changing. This is similar to survey respondents’ indicating that the economy is their most important issue during a recession or stock market crash~\cite{poq18}. 

\subsubsection{Sentiment Extraction}: We use VADER, which has been shown to perform well in SM texts~\cite{vader}, as it takes in account acronyms, emoticons, slang etc, which are ever present in SM. For each post, we calculate the compound sentiment, which is a normalized score between -1 (most extreme negative) and +1 (most extreme positive) and, thus, provides a single measure of sentiment for a given post. We also use LIWC (Linguistic Inquiry and Word Count) to extract features that correspond to emotional, structural and psychometric components present in the text. LIWC classifier categorizes the words in each post into 64 categories related to social, cognitive, personal, informal language etc.

\subsubsection{Comments Metadata}: : We also extract the compound sentiment feature for all comments on each post and then average them to get a single sentiment value. We empirically observe that content features for comments did not better our results, and hence choose not to use them. Further, a qualitative review of comments reveals that most comments do not contain discernible content relevant to markets and the economy. Discarding the content features of comments also helps us to reduce the dimensionality of the resulting feature space.

\subsection{Handling noisy Reddit data}

Learning from Reddit is fraught with problems. Most importantly, the content of posts is corrupted with noise in the form of advertisements, chatter in non-english languages and postings by bots, etc. These mechanisms not only generate random noise, but can artificially inject certain content and sentiment, either by posting links or taking part- in discussions. Since our objective is to sense a signal of human activity for social measurement, it is important that we take into account only the content generated by (human) users. We address these problems in our analysis, by attempting to identify and remove the noisy posts and by focusing only on content generated by “trustworthy” users.
 
To remove noisy posts, we perform clustering on posts embeddings using Kmeans (to get 10 clusters). When the cluster centroids are visualized, we notice that qualitatively different clusters emerge. Some clusters correspond to resume writing, tips on interviewing and changing of jobs while others corresponded to economic conditions, often with political undertones. We find that some of the clusters were related to non-english languages, advertisements and postings by bots etc.; so we remove these clusters for downstream tasks.


To identify content generated by “trustworthy” users, we calculate and assign quality metrics to all individual users, based on their interactions within the Reddit community. Our idea is to assign weights to users that signify our trust in them. Highly trustworthy users get higher weights. We build a logistic regression classifier to distinguish between more and less trustworthy users. We associate trust with varied quantitative features extracted from users attributes and behavior, including their activeness, “recency” in interaction, participatory behavior, karma points over the years etc [10]. To create a training set, we select a subset of users and label them as trustworthy (say label = 1) or not. To attach a label, l to a user u as 1, we retrieve details of users such as whether they served as moderators, whether they participated in discussions over the years with no controversy, and whether they had ever been banned from Reddit. We train a binary logistic regression classifier with an accuracy of 67\% for 10-fold cross-validation. Next, we use this classifier to generate predictions for the test set. Along with these predictions, our classifier also provides probabilistic trustworthiness scores for each user. We calibrate these scores, such that they are distributed exponentially with a long tail, thus ensuring that the distribution of “trustworthy” users in the test set mirrors that in the training set. This calibrated score is the weight of a user.

\subsection{Creating longitudinal covariate data}
Let $p_{t',u}$ denote a post, that is indexed by time $t'$ and authored by user, $u$ and let $s_{p_{t',u}}$ be the score for that user $u$. We denote ${\bf x}_{p_{t',u}}$ as a vector of features extracted from the post $p_{t',u}$. The covariates for a time-interval $t$ (day or month), can be calculated as the weighted average of the features of the posts written during that time-interval, as follows:
\begin{equation}
    {\bf x}_t = \frac{\sum_{\forall t'\in t}{\bf x}_{p_{t',u}}s_{p_{t',u}}}{\sum_{\forall t'\in t}s_{p_{t',u}}}
\end{equation}

Thus, each posts' content and sentiment features are weighted by the corresponding users' weights. We normalize by the sum of user weights for that time-interval to account for the embeddings. Note that we can set $s_{p_{t',u}} = 1$, meaning that user scores are not used.

\subsection{Gaussian Process Regression with Matern Kernel}
We present a Bayesian non-parametric approach to modeling the non-linear relationships between the longitudinal covariates and the regression targets using a Gaussian Process (GP), and is specifically designed for longitudinal polling data. GPs belong to the class of Bayesian non-parametric models, where no assumptions are made on the functional form of the relationships between covariates and targets, and thus, these methods are known to learn highly non-linear boundaries.

A GP is a stochastic process, indexed by ${\bf x} \in \mathbb{R}^d$, and is completely specified by its mean $m({\bf x})$  and its covariance/kernel function $k({\bf x},{\bf x'})$, as shown in: $f({\bf x})  \sim  GP(m({\bf x}),k({\bf x},{\bf x'}))$. The mean is usually assumed to be 0 ($m({\bf x}) = 0$), and covariance between any two evaluations of $f({\bf x})$ is  $k({\bf x},{\bf x}') = \mathbb{E}[(f({\bf x})- m({\bf x}))(f({\bf x}') - m({\bf x}'))]$, where $m({\bf x}) = \mathbb{E}[f({\bf x})]$. The covariance function defines the notion of nearness or similarity in GPs, and thus encodes different types of nonlinear relationships between the covariates and targets. For details of GP for machine learning, the reader is directed to ~\cite{gpml}.

As we are interested in making inferences about the relationship between covariates and targets, i.e. the conditional distribution of the targets given the covariates. We pose our problem as a regression task. GP regression is stated as $y_i  \sim  \mathcal{N}(f({\bf x}_i), \sigma^2_n), \forall i$. Given training data, $\mathcal{D} = \{{\bf x}_i,y_i\}_{i=1}^N$ and a GP prior on $f()$, the posterior distribution of $y_*$ (for an unseen input vector, ${\bf x}_*$), is a Gaussian distribution, with the following mean and variance:
\begin{eqnarray}
  \bar{y}_* := \mathbb{E}[y_*] & = & {\textbf{k}}^\top(K + \sigma^2_nI)^{-1}{\textbf{y}}\label{eqn:gprmean}\\
  \sigma^2_* := \text{var}[y_*] & = & k_{*} - {\textbf{k}}^\top(K + \sigma^2_nI)^{-1}{\textbf{k}} + \sigma^2_n
\label{eqn:gprvar}
\end{eqnarray}


Here, ${\textbf{y}} = [y_1,y_2,\ldots]^\top$, and $K$ is a matrix which contains the covariance function evaluation on each pair of training data, i.e., $K[i,j] = k({\textbf{x}}_i,{\textbf{x}}_j)$, ${\textbf{k}}$ is a vector of the kernel computation between each training data and the test point, i.e., ${\textbf{k}}[i] = k({\textbf{x}_*},{\textbf{x}}_i)$, $k_* = k({\textbf{x}_*},{\textbf{x}_*})$, and $I$ is an identity matrix.


We use the Matern 3/2 class of covariance functions given by:
\begin{eqnarray}
k_{v=3/2} (r) =  (1 + \frac{\sqrt{3}r}{l})exp(- \frac{\sqrt{3}r}{l}) 
\label{eqn:matern}
\end{eqnarray}

where r = |x - x'|. This covariance function is a product of an exponential and a polynomial of order 1, and is suited for learning non-smooth behavior, such as those exhibited by financial time-series~\cite{ghosal}. 



GP regression (GPR) provides a measure of uncertainty along with the predictions via the variance in Equation~\ref{eqn:gprvar}. The uncertainty is a measure of trust provided by our model, and might assist in algorithmic decision making by telling at where (time points) and how much to trust the individual predictions.

\textbf{Training the model}: The learning of hyper parameters in GPR is termed as training the model. The unknown parameters are: the hyper-parameters of the kernel function, $\ell, \sigma_f^2$, and the variance of the error term, $\sigma_n^2$. These are estimated by maximizing the marginalized log-likelihood of the targets in the training data, ${\textbf{y}}$, using conjugate gradient descent~\cite{gpml}.

Once the model is trained, the prediction is then obtained at the inference time-point using ~\eqref{eqn:gprmean} and uncertainty using ~\eqref{eqn:gprvar}.

\begin{algorithm}[htp]
{\footnotesize
  \SetAlgoLined\DontPrintSemicolon
  \setcounter{AlgoLine}{0}
  \SetKwFunction{algo}{GPRM}
  \SetKwProg{myproc}{Procedure}{}{}
  \myproc{\algo{${\bf x},{\bf X}^{Train},{\bf y}^{Train}$}}{
  $\theta \leftarrow {\tt train}({\bf X}^{train},{\bf y}^{train})$ \;
  $\bar{y},\sigma^2 \leftarrow {\tt predict}({\bf X}^{train},{\bf y}^{train},{\bf x},\theta)$\;
  \KwRet{$\hat{y}_t = (\bar{y},\sigma^2)$}\;}{} 
  \vspace{0.2in}
  \setcounter{AlgoLine}{0}
  \SetKwFunction{algo}{Monitor}
  \SetKwProg{myalgo}{Algorithm}{}{}
  \myalgo{\algo{${\bf X}_{0:T},{\bf y}_{0:T},\Delta,w,\alpha$}}{
  $\hat{\bf y}_T \leftarrow []$\;
  \For{$t = (\Delta+w+\alpha):T$}{
    $\bar{y}_t,\sigma^2_t \leftarrow {\tt GPRM}({\bf x}_{t+\alpha},{\bf X}_{t-\Delta-w+\alpha:t-\Delta+\alpha},{\bf y}_{t-\Delta-w:t-\Delta})$\;
    $\hat{\bf y}_T \leftarrow [\hat{\bf y}_T;(\bar{y}_t,\sigma^2_t)]$\;
  }
  \KwRet{$\hat{\bf y}_{0:T}$}\;}{}
  \vspace{0.2in}
  \SetKwFunction{algo}{MonitorWithMissingTargets}
  \SetKwProg{myalgo}{Algorithm}{}{}
  \myalgo{\algo{${\bf X}_{0:T},{\bf y}_{0:T},w$}}{
    \For{$t = (w+1):T$}{
        \If{{\tt isMissing}($y_{t}$)}{
            $\bar{y}_t,\sigma^2_t \leftarrow {\tt GPRM}({\bf x}_t,{\bf X}_{t-w:t-1},{\bf y}_{t-w:t-1})$\;
            $y_{t} \leftarrow \bar{y}_{t}$\;
        }
    }
  \KwRet{${\bf y}_{0:T}$}\;}{}
  \caption{Algorithms for predicting consumer confidence using longitudinal covariates}
  \label{alg:alg1}
  }
\end{algorithm}

\subsection{Computational framework}
\textbf{Notation}: We denote the time series of covariates as ${\bf X}_t \equiv {\bf x}_1,{\bf x}_2,\ldots,{\bf x}_t$, where each ${\bf x}_t \in \mathbb{R}^d$; $d$ refers to the number of features extracted from the Reddit data. 
We denote a time series starting at index $t_1+1$ and ending at index $t_2$ as ${\bf X}_{t_1:t_2}$. The target time series, consisting of the consumer confidence indices, is denoted as ${\bf y}_t \equiv y_1,y_2,\ldots,y_t$, where each $y_i \in \mathbb{R}$.

\textbf{Algorithm details}: We pose the problem of predicting consumer confidence index as a regression task, where the longitudinal covariates are ${\bf X}_t$ and the targets are ${\bf y}_t$. To make prediction at a given time step $t$, $\hat{y}_{t}$, we train the GPR model with Matern kernel, referred to as GPRM in Algorithm~\ref{alg:alg1}, on ${\bf X}^{train}$ and ${\bf y}^{train}$. The training data is the historical data derived from fixed length window of size $w$. The learning in our method is accomplished using data residing in window $w$, which is advantageous as our estimates are governed by only recent relationship between covariates and targets that is encoded within the window.

We introduce the notion of a \textbf{prediction step} (indicated by $\Delta$), which encodes how far ahead can our model reliably predict. $\Delta$ can be 1, in which case the test time point is the next time point. Our training data starts at ${\bf X}_{t-\Delta-w}$ and ends at ${\bf X}_{t-\Delta}$ (inputs) and, similarly, the targets start at ${\bf y}_{t-\Delta-w}$ and ends at ${\bf X}_{t-\Delta}$ (targets). 

We introduce another variable called \textbf{correspondence lag} (denoted as $\alpha$), which is the difference between the start of window for the covariates and targets. It studies how the data generating process for Reddit correspond to survey data - does the signal in Reddit lags or leads the survey data, i.e. if the chatter on Reddit is an after-effect of the consumer sentiment index; or does it precede the market sentiment. 

Introduction of $\alpha$ modifies the extents of the training data as follows: To make predictions at time $t$, the training data starts at ${\bf X}_{t-\Delta-w+\alpha}$ and ends at ${\bf X}_{t-\Delta+\alpha}$ (for inputs) and the targets start at ${\bf y}_{t-\Delta-w+1}$ and ends at ${\bf X}_{t-\Delta}$. 

The training data is used to estimate the kernel hyper-parameters of the model, denoted as $\theta_t$. At inference, the training data along with the input at $t$, ${\bf x}_t$, are used to estimate the target, denoted as $\hat{y}_t$, which is a Gaussian distribution with mean $\bar{y}_t$ and variance $\sigma^2_t$. We use the variance, $\sigma^2_t$, as the model uncertainty at $t$. To make predictions at $t+1$, the training window is shifted to the right by 1, while maintaining its size as $w$. The method to produce estimates for the entire time series is described as the routine {\tt Monitor} in Algorithm~\ref{alg:alg1}. 

\begin{table}
	\centering
	\begin{tabular}{|c|c|c|c|} 
		\hline
		Subreddit       &  Posts  & Comments   &  Users  \\
		\hline
		Economy  &  114040 & 361725 & 16313 \\
		Finance    & 198176 & 255006 & 53680 \\
		Jobs &  267767 & 1384548 & 125422 \\
		\hline
	\end{tabular}
	\caption{Reddit data description. Data was available from March 2008 until August 2019 for all the subreddits.}
	\label{tab:datadesc}
\end{table}



The {\tt MonitoringWithMissingTargets} routine in Algorithm~\ref{alg:alg1} describes how our framework takes care of the scenario when we want to reduce the frequency of survey data i.e. some of the $y_{t}$ values are not available. To fill in the missing value at $t'$, we use the historical data till $t'-1$ for training GPRM model and predictions are derived at $t'$. The mean of the predictions $\bar{y}_{t'}$ is now our estimates at $t'$. This procedure is repeated for all the unavailable target values. 

\begin{figure*}
    \includegraphics[width=0.23\textwidth]{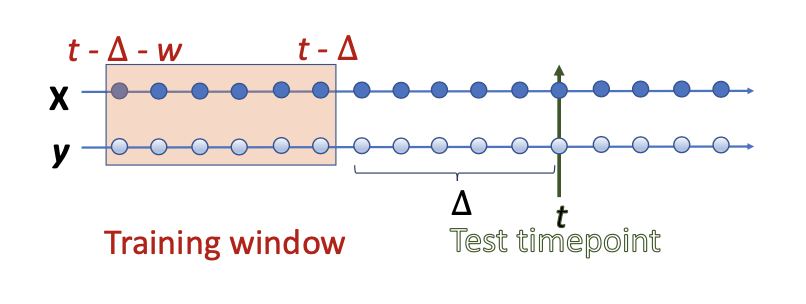}
    \includegraphics[width=0.23\textwidth]{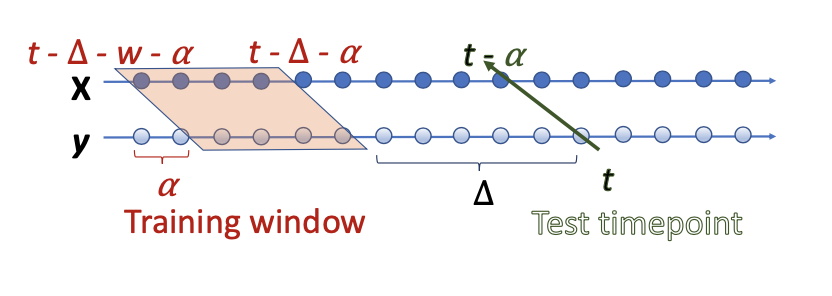}
    \includegraphics[width=0.23\textwidth]{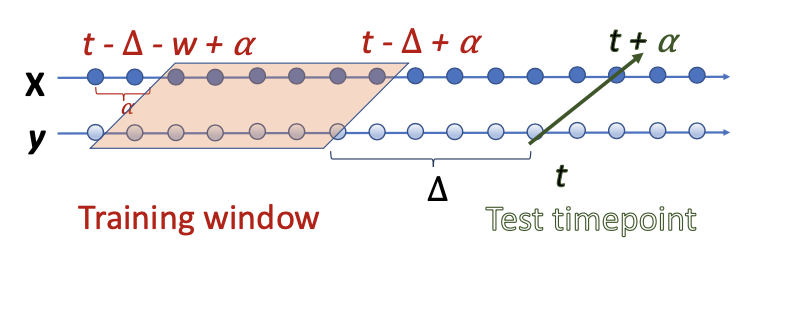}
    \includegraphics[width=0.23\textwidth]{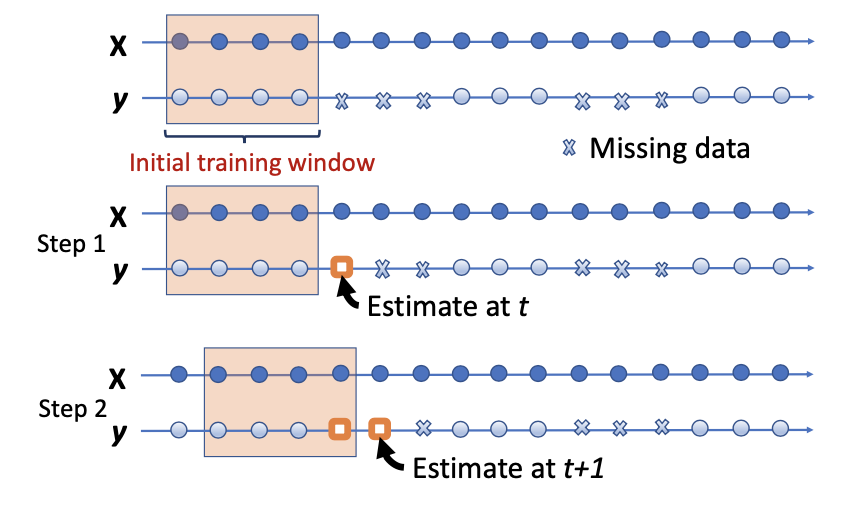}
    \caption{\footnotesize{Pictorial depiction of our experimental strategy. X depicts the covariates, Y the targets. For simplicity X and Y are shown as a line with the dots depicting each time point, $\delta$ is the prediction step and $\alpha$ is the correspondence lag. The orange color window of size $w$ depicts the training window. The green arrow marks the test point at $t$. From L to R - In the leftmost figure, the training window is a rectangle, meaning the X and Y span the same time periods. The next two figures show the scenario where there is a correspondence lag ($\alpha$) between X and Y. The rightmost diagram shows our testing strategy for reducing the frequency of surveys. Step 1 shows how missing survey data is estimated at time $t$, using historical data at $t-1$. These estimates are used to fill in missing data at subsequent times.}}
    \label{fig:testing}
\end{figure*}

\section{Experiment Setup}
\textbf{Reddit data}: We focus on the economy, jobs and finance subreddits because these contain  chatter we expect to be related to consumer confidence. An alternative to choosing subreddits is to conduct a keyword search of all Reddit content for a set of chosen keywords. However, we rejected this method because it is difficult to know whether one has chosen the ’right’ and ’complete’ set of keywords. Reddit data was extracted using Google BigQuery\footnote{\url{https://console.cloud.google.com/bigquery}}. Table~\ref{tab:datadesc} describes the time extents and characteristics of the three relevant subreddits. Our analysis indicated that user features did not improve our performance in these experiments, so they are not included in the reported results.

The data spanned from March 2008 until August 2019 for all the subreddits. The raw covariates consisted of 4 sentiment features extracted using VADER, 64 sentiment features extracted using LiWC, 100 content features extracted using {\em doc2vec}, and 1 sentiment features extract from the comments corresponding to a post. The dimensionality of the LiWC and content features was reduced using {\em Principal Component Analysis} (PCA) by using top principal components that account for 90\% of covariance in data. 


\textbf{Consumer Confidence Data}: The {\em Index of Consumer Sentiment} (ICS) collected by the Institute for Social Research at the  from University of Michigan is an important survey that measures U.S. consumer confidence. Researchers have extensively studied it in social sciences, as well as in prior works on SM forecasting. The survey consists of responses aggregated to a single index of five questions based on an individual's personal finances, their outlook for the economy in near future and their recent buying decisions. A nationally representative sample of approximately $500$ respondents is drawn, and the poll is administered via telephone interviews. Although this indicator is released at monthly, the daily time-series is also available. We perform our analysis at both daily and monthly granularity. While the monthly ICS data is available from 1978, the daily ICS data is only available from January 2008 to May 2017. It is important to note that there are error bounds corresponding to target data too. With monthly ICS, the 95\% confidence interval is +/- 3.29\%, while the confidence interval with daily data is not available to us.

\textbf{Smoothing and Window Sizes}: 
For daily analysis, the time span between March 2008 and May 2017 was considered. For monthly analysis, time span between March 2008 and August 2019 was considered. Both the covariate and daily data were smoothed using moving average with 7, 14, 28-day window lengths. The decision to smooth was taken to account for large variation in daily time-series and to make trends apparent. Also, smoothing allowed us to experiment with different values of prediction steps and correspondence lags into our model. The window size($w$) is set to $48$ for monthly data and $730$ for daily data.


\textbf{Performance Metrics}: For evaluation, the prediction at point t, $\hat{y}_t$ is compared with target $y_t$, and the error is calculated. Model performance is measured using mean absolute error (MAE), which is the average of the absolute errors for a set of predictions and root mean squared error (RMSE), which is the square root of the mean absolute errors. Both MAE and RMSE are in the same units as the target variable, and their lower values indicate better fit. Both are measures are frequently used to gauging the fit between the predictions of a regression model and  real-world target data.

We study the correlations of the predictions versus the survey data time series using Pearson's correlation. Correlations with smoothed and autocorrelated time-series data can lead to amplified and spurious values of Pearson’s correlations, especially when outliers are present~\cite{dcca}. Therefore, we  also report detrended cross-correlation analysis (DCCA). DCCA is known to be a robust measure of cross-correlations between two different but equal length time series, in the presence of non-stationarity. Our method provides Bayesian uncertainty with each prediction, which is the regression model's predictive variance. Less uncertainty with the predictions indicates greater confidence in our model’s predictions.

\section{Experiment Design}
Experiments were conducted to assess three important characteristics of our model, described as follows:

\textbf{1. Performance accuracy of the model}: Experiments with prediction steps, $\Delta$, test how our model's accuracy varies when it is used to predict at future windows, say 7, 14, 28 days ahead. We also experiment with correspondence lag, $\alpha$, to study how the data generating process for Reddit correspond with that of the survey data. 

\textbf{2. Reducing frequency of surveys}: We investigate the extent to which Reddit data can  reduce the need for frequent surveys, while preserving a high degree of correlation with traditional survey measures. Our approach is to estimate consumer confidence with reduced survey availability (e.g., survey data is available every other month or after two-months, and so on), and then use the daily Reddit data to estimate the ICS values for the months when survey was missing. In this way, we can estimate how far in the future the model can reliably estimate the ICS values before its performance drops significantly. Our experimental setup is as follows: In first scenario, we "remove" every alternate 28-day period of ICS data, after an initial training window of 2 years. The algorithm {\tt MonitorWithMissingTargets} is used to estimate the ICS values for the missing days, one day at a time. After each day, the estimated value for that day is ``fed back'' into the training data for the subsequent prediction. Thus, in the following time steps, the model uses a mix of observed and estimated target values to produce the estimates. This process is repeated until the next survey data is available. The {\tt MonitorWithMissingTargets} algorithm is applied again whenever there are subsequent unavailable survey targets. The estimated targets are compared with the ground truth values using the different performance metrics discussed above. The same experimental setup is repeated for scenarios in which the survey data is available at a further reduced frequency, i.e., it is available for a 28 day period after every 56 days, then 84 days, and 112 days. Hence, for each window, the model must generate predictions using fewer days of survey data. 

\textbf{3. Baseline Methods}: It is important to put these research methods (ones that use alternative data sources for prediction) in context by producing a baseline to establish their real added value, and will put these methods into practice. In order to evaluate the practical utility of substituting predictions based on SM data for frequent, traditional polling, we compare the accuracy of our estimates of consumer confidence with two baseline models. The first model, which we call “time-only,”  exploits the temporal auto correlation of the targets. The second model uses a single feature, e.g., the sentiment score or content embeddings, to predict the survey targets. The time-only baseline method uses a non-linear weighted averaging of past targets and uses no social media data. It is calculated using our GPR model with only time as the covariate. This baseline model allows us to answer the question: “How accurately can we predict future values of survey data using only past values?”
The single feature based baseline, in contrast, allows us to ask: What is the predictive performance with a single feature (like sentiment score or content score) extracted from SM data?  This model elucidates the net gain achieved by using a more diverse feature space. To answer this, we use two variants of our model - one with a single sentiment covariate, namely the compound score of each post; and the other with principal content score as a covariate.  In both models, the targets are the daily CCI survey data.

\section{Results and Discussion}
We describe the results targeting the three aspects of our experiments:
\begin{table}[thbp]
	\centering
	\begin{tabular}{|c|c|c|c|c|c|}	
	\hline
          Step (months) &   RMSE &  MAE  & Corr &  DCCA   &  Variance \\
          \hline
        	1	&	5.70 &	4.65 &	0.85 &	0.65 &	17.00 \\
        	2	&	5.90 &	4.83 &	0.85 &	0.59 &	18.35 \\
        	3	&	6.20 &	5.19 &	0.85 &	0.54 &	19.06 \\
            4	&	6.47 &	5.37 &	0.84 &	0.50 &	19.89 \\
            5	&	6.90 &	5.60 &	0.83 &	0.50 &	21.30 \\
     \hline
	\end{tabular}
	\caption{{Performance of the model using monthly data. Step size refers to the number of months in advance at model predictions are calculated ($\Delta$).}}
	\label{tab:accmonthly}
\end{table}			

\begin{figure*}[htbp]
    \centering
    \includegraphics[width=0.8\textwidth]{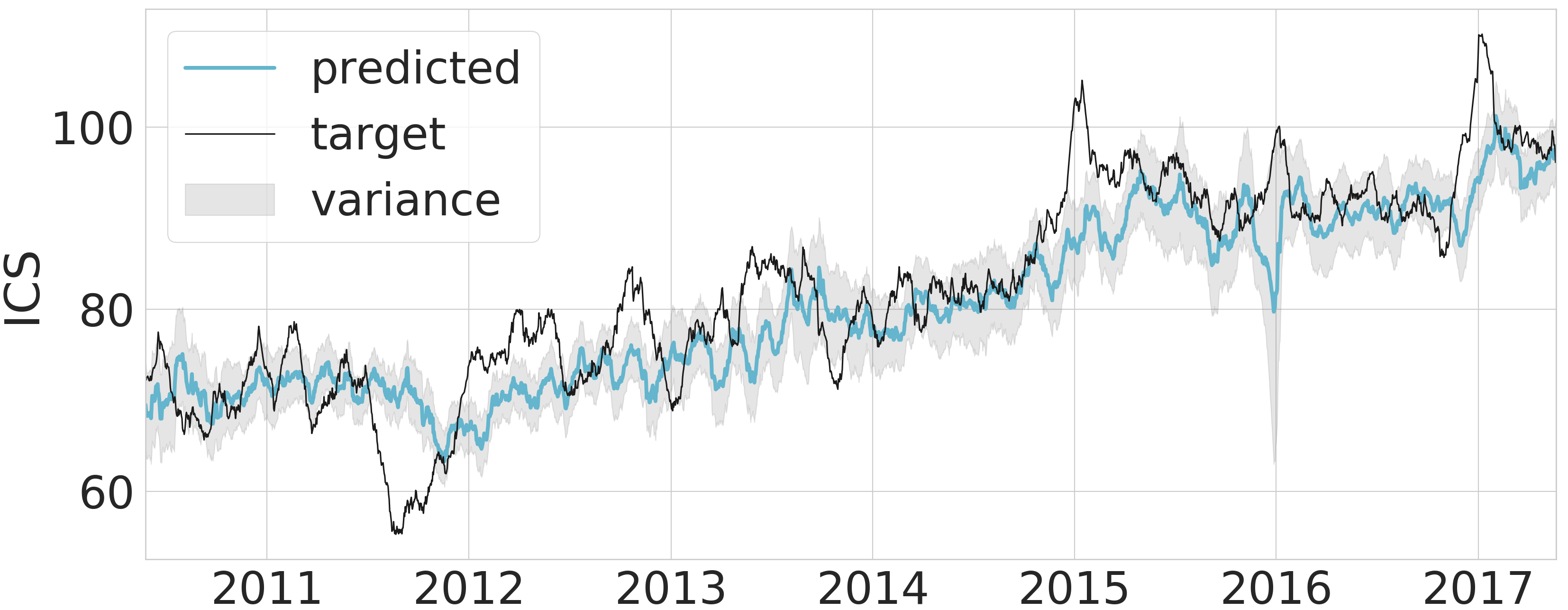}
    \caption{\footnotesize{Comparing the model predictions with observed targets at daily granularity (smoothing window size - 28, $\Delta$ - 28)}}
    \label{fig:accdaily}
\end{figure*}

\begin{figure}[htbp]
    \centering
    \includegraphics[width=0.47\textwidth]{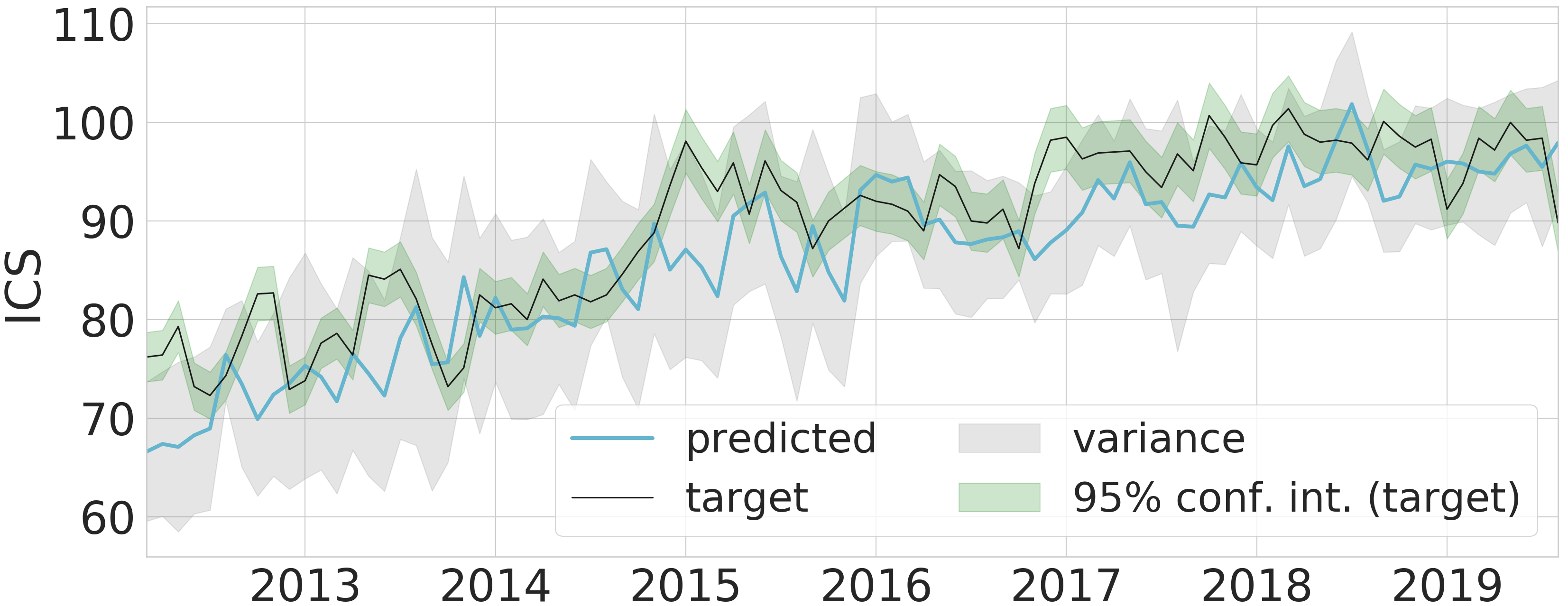}
    \caption{{Model performance using monthly target data. The green envelope around the target time-series signify the 95\% confidence interval (which is +/- 3.29)}}
    \label{fig:accmonth}
\end{figure}
%

\textbf{Performance accuracy of the model}: Figure~\ref{fig:accdaily} describes how the predicted time-series compares with target ICS values for daily data with prediction step ($\Delta$) as 28 days. It shows that our model based on Reddit data does capture the overall trend in ICS values, and also some of the finer longitudinal movements, like those between 2012 and 2015. Our model under-predicts for some of the noticeable peaks in Jan. 2015 and Jan 2017. We also observe that our model does not perfectly capture the significant drop in ICS during later half of 2011, but it does predict a noticeable dip around that time point, albeit with a delay. A noticeable mismatch between the targets and our predictions occur at the beginning of 2016. We attribute it likely to significant underlying policy changes in the Reddit platform around this time~\cite{shadowban}. 
However, this prediction is also marked with large value of uncertainty given by our model. 
Figure~\ref{fig:accmonth} describes how our predictions track the macro trends, as well the micro-movements in the time-series at monthly granularity, with $\Delta$ = 1. It is important to notice that most of the time the ICS values lie within the uncertainty bounds of the predictions. 

Table~\ref{tab:smooth} details the quantitative performance of the model for varied smoothing and prediction steps. When focusing on the RMSE and MAE errors, we notice that the model works well in predicting the ICS targets for even 28 days in advance. For daily granularity, when $\Delta$ is close to 1, the model was able to predict the targets almost perfectly, most likely because of the presence of strong auto-correlation in the smoothed target time-series. To avoid such spurious results, we do a stricter test by measuring the performance for $\Delta \geq 28$. We notice that more smoothing tends to capture consistent trends in the longitudinal timeseries.

Table~\ref{tab:lag} details how the model behaves when $\Delta > 28$). As $\Delta$ increases, the performance of our model degrades but not very rapidly. The trend is consistent with different smoothing values, and across all measures of performance. 

The experiments performed with correspondence lag - $\alpha$ did not point to either Reddit data lagging or leading the survey variables, and the best results were obtained when no correspondence lag is input to the model. This states that the data generating processes of Reddit occurs simultaneously with the survey data, and thus for Reddit it is best to temporally align it with survey data for the same month.


\begin{table}[htbp]
	\centering
	\begin{tabular}{|c|c|c|c|c|c|c|} 
		\hline
	    Smoothing    &   $\Delta$  &  RMSE &  MAE  & Corr & DCCA   &  Variance \\
		\hline
		28	& 28	&	5.52 &	4.23 &	0.89 &	0.78 &	4.39 \\
	    28	& 42	&	5.96 &	4.52 &	0.87 &	0.73 &	4.66 \\
		28	& 56	&	6.31 &	4.75 &	0.85 &	0.67 &	4.83 \\
		28	& 70	&	6.53 &  4.92 &  0.85  &	0.62 &  4.95 \\
		28 & 168    &   7.79 &  6.21 &  0.81  &  0.36 & 5.63 \\
		\hline
	\end{tabular}
		\caption{{Model performance on daily data for different prediction step sizes, $\Delta$. Smoothing window size - 28.}}
		\label{tab:lag}
\end{table}	
\textbf{Reducing frequency of surveys}: Table~\ref{tab:calibrate} shows that Reddit data can be used to reduce the cost of surveys, by estimating accurate responses in periods when no survey was undertaken. The step in the Table~\ref{tab:calibrate} refers to the frequency of conducting the survey - Step of 2 means survey can be done every $2^{nd}$ month; step of 3 means every $3^{rd}$ month, so on. As we decrease the frequency of the surveys, the estimates during the off-month period decrease at a steady rate. 
\begin{table}[htbp]
	\centering
	\begin{tabular}{|c|c|c|c|c|c|c|} 
		\hline
	    Smoothing    &  $\Delta$  &  RMSE &  MAE  & Corr. & DCCA   &  Variance  \\
	    \hline
		7 & 28 & 7.30 &	5.68 & 0.80	& 0.71 & 14.19 \\
		14 & 28 & 6.32 & 4.91 & 0.85 & 0.74	& 7.60 \\
		28	& 28 &	5.52 &	4.23 &	0.89 &	0.78 &	4.39 \\
		\hline 
	\end{tabular}
	\caption{{Model performance on daily data for different smoothing window sizes ($\Delta$ = 28).}}
	\label{tab:smooth}
\end{table}	

\begin{figure*}[htbp]
    \centering
    \includegraphics[width=0.8\textwidth]{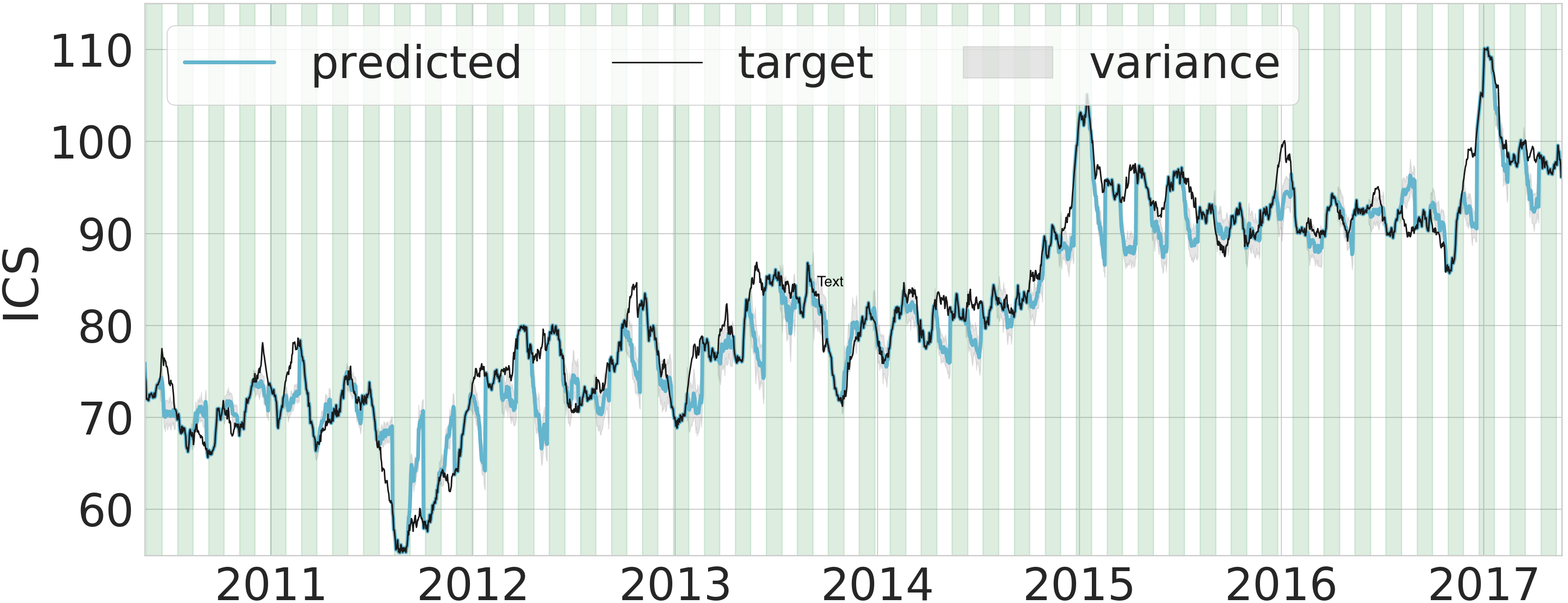}
    \caption{\footnotesize{Comparison of our predictions with survey data for experiments conducted to reduce the frequency of surveys by half. The shaded green regions marks the 28 day periods when surveys are available, which is every alternate time-period. Notice how our estimates closely follow the targets. A close-up of a random sample of this time-series is shown in Figure 5 (leftmost).}}
    \label{fig:calibrate2}
\end{figure*}

\begin{figure*}[htbp]
    \centering
    \includegraphics[width=0.33\textwidth]{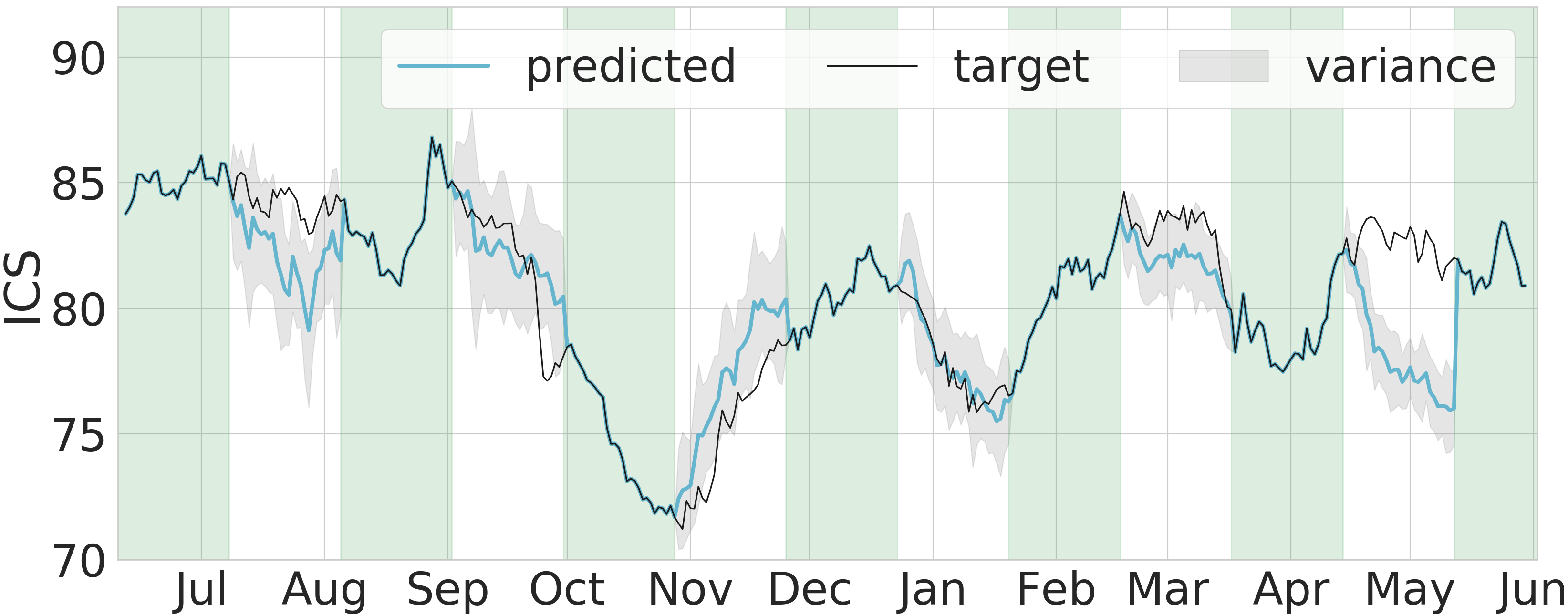}
   \includegraphics[width=0.33\textwidth]{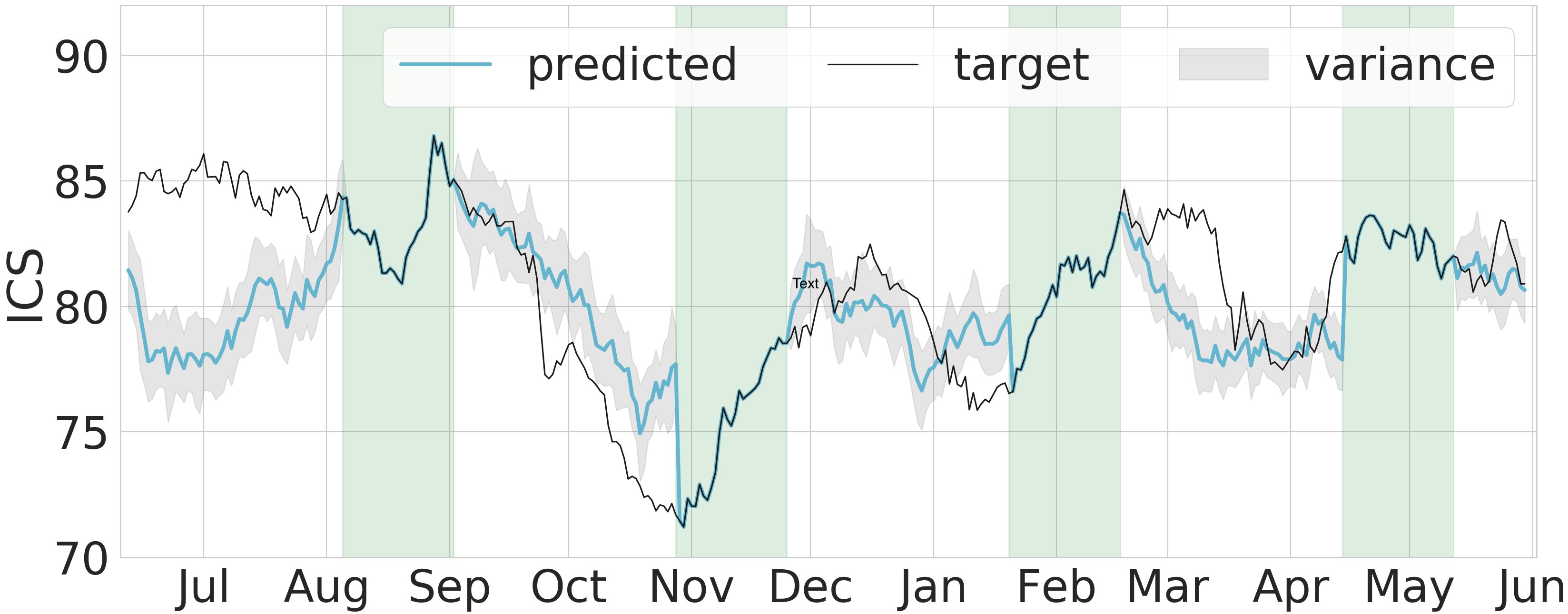}
    \includegraphics[width=0.33\textwidth]{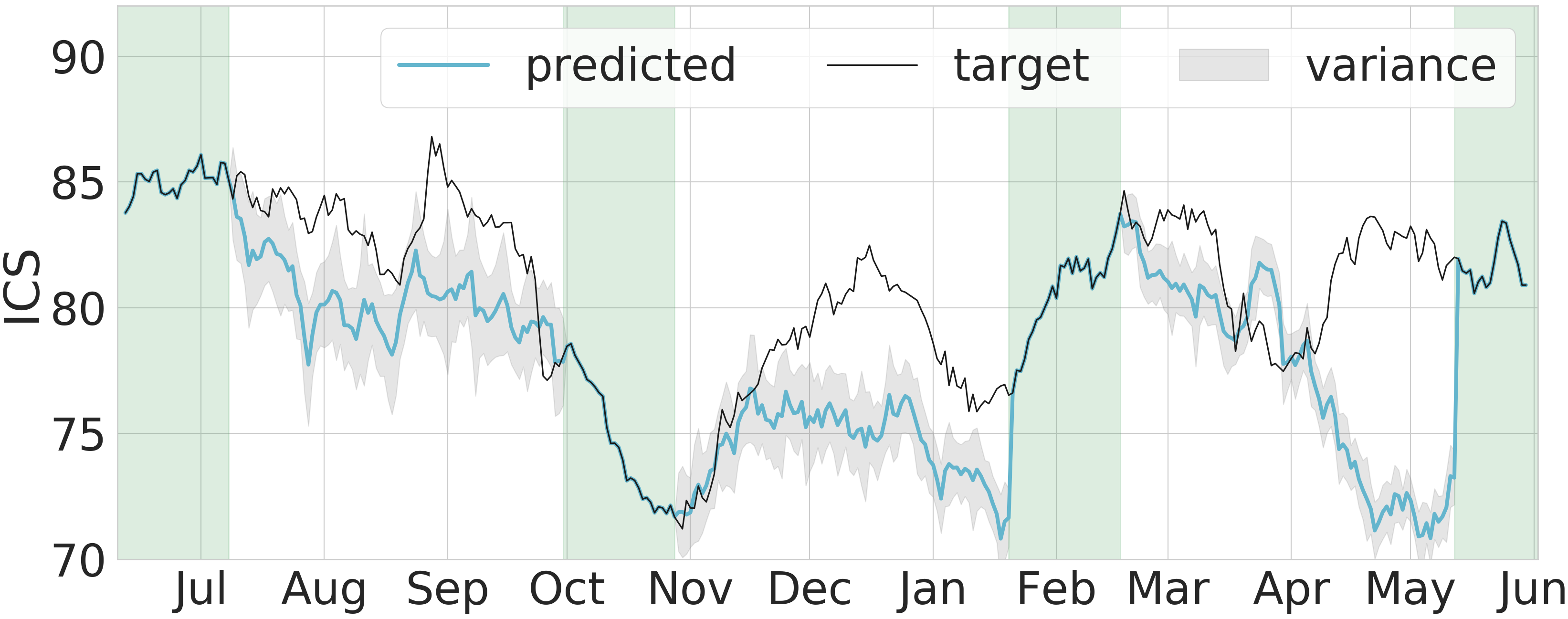}
    \caption{\footnotesize{A snapshot of the time-series (for the year 2014) displaying targets, predictions and corresponding uncertainties for the set of experiments to reduce the survey frequency. The shaded green color marks the 28 day periods when surveys are available. The white regions mark the time period when our model was used to estimate survey values. Our predictions are shown as blue. It is important to see how the model estimates and uncertainties compare with the targets at these time-periods. From L to R - surveys available for (leftmost) every alternate 28 day period, (middle) for every third 28 day period and (rightmost) surveys available for every fourth 28 day period. As we go from left to right, the frequency of surveys decreases. Notice how uncertainty varies with the predictions in the off-months when survey is not done.}}
    \label{fig:cali}
\end{figure*}

Figure~\ref{fig:calibrate2} shows the performance of our model when survey data for alternate month was made unavailable. We see how our predictions closely track the targets, and is an encouraging result showing that Reddit data can to used to supplement surveys. In figure~\ref{fig:cali}, we take a snapshot the timeseries (July-Jun 2014), so that we focus on how our predictions and their associated uncertainties vary when surveys are run every second, third and fourth month respectively for the duration of our experimental data. While the estimates are close to the target for some of the months, they are off for others. 

Comparing the left to the rightmost diagram in Figure~\ref{fig:cali}), we notice that as more time elapses between surveys, the quality of the estimates suffer. This is because to make prediction at time $t$, the model relies on a window of past 2 years of data, where the missing survey points have been replaced by the model estimates. When the time elapsed between the surveys is large, the predictions rely mostly on model estimates. However, once the next survey is available, the model can again reliably predict the subsequent target values.

\begin{table}[htbp]
	\centering
	\begin{tabular}{|c|c|c|c|c|c|c|} 
		\hline
    	Smooth   &   Step  &  RMSE &  MAE  & Corr & DCCA   &  Variance  \\	\hline
		28	 & 2 &	4.03 &	3.11 &	0.94 &	0.98 &	0.75 \\
		28	& 3	 & 5.51	& 4.05	& 0.90 & 0.93 &	0.63 \\
		28	& 4	 & 7.24	& 5.70	& 0.84 & 0.77 &	0.84 \\
		28	& 5	& 7.45	& 5.70	& 0.85 & 0.71 & 0.46 \\	 	
       \hline
	\end{tabular}
	\caption{{Experiments to reduce the survey frequency using Reddit data for every second, third and fourth month.}}
	\label{tab:calibrate}
\end{table}			
\textbf{Baseline Methods}: The time-only baseline produces an RMSE of 12.43 and MAE of 9.75; which is significantly higher than the corresponding results obtained with the Reddit data. Similar performance degradations were observed for other metrics. This bolsters the fact that using Reddit data does help to make better predictions for consumer confidence index, than just using past target values. 

%
%

Table~\ref{tab:baseline} details the advantage of combining both sentiment and content based features to achieve higher performance gains. 

\subsection{Limitations}
Despite their advantages, sensing with social media data has some pitfalls.  Some of the most important limitations include the lack of representativeness of SM users compared to the general population and, the lack of demographic information present in social media that could help rectify the bias. Also, when researchers use social media data they either use a data dump (e.g. Reddit), or may be provided a sample of the data (e.g. Twitter). In both cases, studies based on these datasets suffer from sampling bias. Keeping the limitations in mind, we suggest that Reddit data can be used to supplement  traditional survey responses to consumer confidence, especially for near future; but may not completely replace these surveys.



\section{Conclusions and Future work}
We show how using a diverse set of longitudinal data extracted from Reddit can model the relationship of these time-varying covariates with the polling data (i.e. consumer sentiment index). Our modeling framework is based on Gaussian Process Regression, which produces uncertainty estimates along with the predictions. Using a robust testing strategy, we show how our methodology can be helpful in reducing the frequency of surveys.

As future steps, we are working to incorporate Twitter data into our model, and expand our methodology to other survey data in order to better understand  the conditions under which information extracted from social media can substitute for or compliment existing polling methods to gauge public opinion.

\begin{table}[htbp]
	\centering
	\begin{tabular}{|c|c|c|c|c|c|c|} 
		\hline
    	Features     &  RMSE &  MAE  & Corr & DCCA   &  Variance  \\	\hline
        Sentiment Only  &	9.26 &	6.83 &	0.69 &	0.45 & 44.04 \\
        Content Only	&	5.58 &	4.26 &	0.88 &	0.77 &	4.29 \\
        Reddit (all)    &	5.52 &	4.23 &	0.89 &	0.78 &	4.39  \\
        \hline
	\end{tabular}
	\caption{{Impact of adding diverse feature spaces.}}
	\label{tab:baseline} 
\end{table}	

\section*{Acknowledgment}
We gratefully acknowledge the financial support of the cybersecurity cluster programmatic fund from the Institute for Security, Technology and Society (ISTS), Dartmouth College, NH.

\bibliographystyle{abbrv}
\bibliography{acaif}
\end{document}